# Deliberation favours social efficiency by making people disregard their relative shares: Evidence from US and India

Valerio Capraro[1,2], Brice Corgnet[3], Antonio M. Espín[2*], Roberto Hernán-González[4]

1. Center for Mathematics and Computer Science (CWI), Amsterdam, 1098 XG, The Netherlands.
2. Department of Economics, Middlesex University Business School, Hendon Campus, The Burroughs, London NW4 4BT, UK.
3. EMLYON Business School, Univ Lyon, GATE L-SE UMR 5824, F-69131 Ecully, France.
4. Business School, University of Nottingham, Jubilee Campus, Nottingham, NG8 1BB, UK.

*Corresponding author: a.espin@mdx.ac.uk

**Abstract**

Groups make decisions on both the production and the distribution of resources. These decisions typically involve a tension between increasing the total level of group resources (i.e. social efficiency) and distributing these resources among group members (i.e. individuals' relative shares). This is the case because the redistribution process may destroy part of the resources, thus resulting in socially inefficient allocations. Here we apply a dual-process approach to understand the cognitive underpinnings of this fundamental tension. We conducted a set of experiments to examine the extent to which different allocation decisions respond to intuition or deliberation. In a newly developed approach, we assess intuition and deliberation at both the trait level (using the Cognitive Reflection Test, henceforth CRT) and the state level (through the experimental manipulation of response times). To test for robustness, experiments were conducted in two countries: the US and India. Despite absolute level differences across countries, in both locations we show that: (i) time pressure and low CRT scores are associated with individuals' concerns for their relative shares; (ii) time delay and high CRT scores are associated with individuals' concerns for social efficiency. These findings demonstrate that deliberation favours social efficiency by overriding individuals' intuitive tendency to focus on relative shares.

**Keywords:** efficiency, equality, dual process models, intuition, deliberation

# Introduction

Groups of individuals, from small-scale societies to large modern organizations, are typically involved in both the production and the distribution of resources (1, 2). Because the distribution process may result in the destruction of part of the resources, there often exists a fundamental conflict between the concern for total group resources (i.e. "social efficiency") and the concern for group members' relative shares of the group resources.

The conflict between equality and efficiency has indeed traditionally been at the center of the debate in distributive justice and social choice theory (1-4). Less attention has been paid, however, to "antisocial" concerns such as spitefulness which, like concerns for equality (egalitarianism), also relate to individuals' relative payoffs. Egalitarianism refers to a motivation for reducing payoff differences among individuals whereas spitefulness refers to an individual's willingness to maximize the difference between her own payoff and that of others (5-7). Thus, both egalitarian and spiteful motives may lead an individual to actively change the group members' relative shares even if the resulting distribution wastes resources and is thus socially inefficient. Efficiency, egalitarian and spiteful motives may not only conflict with each other but also with self-interest. Yet, people are frequently willing to forego personal gain in order to increase group resources, equalize payoffs or maximize their relative share.

When faced with allocation decisions in which conflicts between social motives may arise, different individuals often act according to different social preferences (5, 8-15). However, much is yet to be learned on the origin of these individual differences in social preferences and on whether they can be exogenously manipulated. Our study aims at answering these questions following a dual-process approach.

Dual-process theories assume that human decisions result from the interaction between two cognitive systems, one that is fast, intuitive, and relatively effortless, and one that is slow, deliberative, and relatively effortful (i.e. the so-called systems 1 and 2 (16-19)). The use of a dual-process lens raises the following general question: given a decision conflict, which option is favoured by the intuitive system? Which one is favoured by the deliberative system? Classifying social decisions as intuitive or deliberative is fundamental for our understanding of human nature. From a practical viewpoint, this will also allow us to design institutions that encourage certain social behaviours and discourage others (20, 21).

Regarding our research question, there is evidence that equality concerns are associated to intuitive emotional processing (4, 22, 23) and that deliberation promotes utilitarian choices that favour "social efficiency" (e.g. save five lives at the expense of one) in moral dilemmas (24-28). In addition, recent trait-level research conducted in laboratory settings in the US and Spain shows that individuals with a more intuitive cognitive style are more likely to choose options that either equalize payoffs between themselves and others (i.e. egalitarian choices) or maximize their own payoff relative to their counterparts (i.e. spiteful choices); by contrast, a more deliberative cognitive style is related to choices that increase the counterparts' payoffs at a *very low* cost for the decision maker thus promoting social efficiency (12, 29). The reported effects have been shown to be robust to controlling for cognitive confounding factors such general intelligence

(12). Relatedly, in contest experiments, more intuitive individuals have been found be more willing to "spitefully" overbid in order to outcompete their counterparts (30).

Based upon this evidence, we hypothesized that when faced with social allocation decisions, people's first impulse is to care about the relative share each individual gets (in either an egalitarian or spiteful manner) whereas deliberation helps override this tendency and preserve social efficiency. Our hypothesis is thus that decisions which rely on intuition are more likely to be driven by the consideration of people's relative payoffs and less likely to be driven by social efficiency concerns. By contrast, deliberative choices are more likely to disregard relative payoffs in favour of social efficiency. In this paper, we test this hypothesis by adopting a novel approach that captures the effect of intuition and deliberation on individuals' social choices at both the trait and the state level. Moreover, to check for robustness, we gathered data from two countries: the US and India.

Specifically, we design an online experiment in which participants from the US and India are asked to make a series of six simple, cognitively undemanding decisions about real monetary allocations between themselves and another anonymous participant (12, 31). Looking at individuals' consistency across decisions, we can classify their choices into three categories of social preferences (5): (social) efficiency, egalitarian and spiteful. Social efficiency refers to a preference for maximising the sum of both individuals' payoffs, whereas egalitarianism refers to a preference for minimising payoff differences between the two individuals. Finally, spitefulness refers to a preference for maximising the decision maker's relative standing. For each category, we use two alternative definitions: one "model-based" definition, based on a *generalized* version of the Fehr & Schmidt (9) model of social preferences; and one "choice-based" definition, based on the number of choices which are consistent with a particular preference. These two approaches to classify people into behavioural types have been extensively used in economics and social psychology, respectively. In addition to these social motives, we also consider self-interest (i.e. the preference for maximising one's own absolute payoff with disregard for others) as an essential motivation when dealing with material resources. For self-interest, both definitions result in the same classification of participants (see Methods for further details).

For the assessment of the role of intuitive vs. deliberative systems in decision making we adopt two strategies. On the one hand, we conducted a *trait-level* analysis by comparing the distribution of social motives between subjects who score low on an updated version of the extended Cognitive Reflection Test (CRT) (32, 33) and those who score high. The CRT consists of a set of questions that all have an intuitive, yet incorrect, answer that should be first ignored to be able to obtain the correct answer. Thus, CRT scores provide a measure of people's ability to suppress automatic/intuitive responses in favour of reflective/deliberative ones. Since answering correctly the CRT requires basic numerical ability apart from reflection, we added a Numeracy Test in order to account for this possible confound (34, 35). On the other hand, we conducted a *state-level* analysis by manipulating participants' cognitive mode using time constraints. Specifically, previous research has argued that time pressure makes people more likely to rely on intuitions (17, 36, 37). By comparing subjects forced to decide in less than 5 seconds (i.e. *time pressure* condition) with those forced to stop and think through their decision for at least 15 seconds (i.e. *time delay* condition) we could (a) further support the results of the trait-level correlational analysis, and (b) establish a causal link between cognitive reflection and social motives (see Methods).

As mentioned, our experiments were conducted using populations from the US and India. Previous research suggests that good institutions can foster social norms that spill over to citizens' everyday behaviour (38, 39). Since the US and India score very differently in corruption indices (40, 41), one may expect that residents in these two countries have developed different preferences. Indeed, behavioural studies show that residents in India are less cooperative (42) and more spiteful (43) than residents in the US. Thus, these two locations represent interesting robustness checks.

# Results

### *CRT and Social Motives*
For the trait-level analysis we assess subjects' cognitive style, intuitive vs. deliberative, using the CRT and study their decisions when there is no time restriction for decision making, i.e. the *neutral* condition (US, n=116; India, n=76). Since in the two "non-neutral" conditions the CRT was performed after the treatment manipulations and the effect of CRT is expected to be milder when time responses are manipulated, the neutral condition is the proper scenario to analyse the effect of CRT on social motives (see next paragraph). In panel A of Figures 1 to 4, we display the proportion of subjects whose choices can be classified according to the aforementioned four categories – social efficiency, egalitarianism, spitefulness and self-interest, respectively –, broken down into below- and above-median CRT scores. For the sake of graphical illustration, the figures are based on above- vs. below-median CRT, whereas the statistical analysis uses CRT score (ranging from 0 to 7) as explanatory variable. The size of the effect represented graphically thus does not directly compare to the size of the effect in the regression analyses, which moreover also control for age and gender as potential confounding factors (32, 44).

We find that the relationship between CRT scores and social motives is substantial and remarkably similar across countries with the exception of the choice-based egalitarian measure. Our regression analysis indeed shows that, for either definition, CRT score is a significant (or marginally significant) predictor of all the categories (Probit regressions with robust standard errors; see Panel A in Tables S1 to S4 in the Supplementary Information (SI)) and the interaction between country and CRT is only marginally significant for the choice-based egalitarian variable (p=0.06; all the remaining p's>0.15; see panel A in Tables S5 to S8). Specifically, higher CRT scores predict a significantly lower likelihood of being classified as egalitarian and spiteful (all p's<0.02), but a higher likelihood of belonging to the social efficiency (both p's<0.01) and self-interest categories (p=0.07). Regarding the only variable where the effect of CRT marginally differs across countries, i.e. choice-based egalitarianism, a joint-significance Wald test on the interaction coefficients reveals that the relationship is significant for the US (p<0.01) but not for India (p=0.56).

We conducted our trait-level analysis using only the neutral-condition sample for two reasons. First, in the two "non-neutral" conditions the CRT was performed after the treatment manipulations. CRT scores can thus be somehow contaminated by spillover effects. Second, in line with dual-process theory the effect of CRT should be milder when either typically-deliberative, high-CRT individuals are forced to choose quickly or typically-intuitive, low-CRT individuals are forced to stop and reflect (that is, in the time pressure and time delay conditions, respectively). Nevertheless, we report the main regression results considering participants in the time pressure and time delay conditions separately (see Tables S9 and S10). We observe that while the sign of the CRT effect does not change in any regression with respect to what we observe in Panel A of Tables S1 to S4 (i.e., in the neutral condition), indeed, the magnitude of the CRT effect is generally reduced, especially in the time pressure condition.

From Panel A in Tables S1 to S4, we observe some significant differences between countries. In particular, residents in India are less likely to be classified as "socially efficient" than residents in the US ($p=0.07$ and $p=0.03$ for the model-based and choice-based definitions, respectively). In the case of egalitarianism, the model-based definition yields a marginally significant difference ($p=0.06$) but the choice-based one does not ($p=0.93$). Moreover, the coefficients of the country variable are of opposite sign in the two regressions. Therefore, we treat the difference on egalitarianism with caution. Regarding spitefulness, in line with Fehr et al. (43), we find that residents in India are significantly more spiteful than residents in the US according to the choice-based definition ($p<0.01$), although not significantly so according to the model-based definition ($p=0.33$; note that the likelihood of finding a significant difference might have been reduced due to the fact that the model-based definition only classifies 9% of subjects as spiteful).

Importantly, when including both numeracy skills and CRT scores as predictors, numeracy is significant in only one out of seven cases, i.e. choice-based social efficiency ($p=0.03$; all remaining $p$'s$>0.11$; see Table S11), indicating that numeracy is unlikely to act as a mediator in the relationship between CRT and social motives. In contrast, CRT remains significant in all ($p$'s$<0.04$) but one regression. The only exception is the model-based spitefulness category, in which CRT turns non-significant ($p=0.33$). Yet, using the choice-based definition of spitefulness, the significant effect of CRT is robust to controlling for numeracy. Thus the effect of CRT on social motives seems to be related to trait reflectiveness and not to numeracy skills.

Therefore, we conclude that, across countries, high cognitive reflection is characteristic of those individuals motivated by social efficiency and, to a lesser extent, by self-interest, but uncharacteristic of individuals whose choices reflect either egalitarian or spiteful motives. These results are thus consistent with previous findings showing that individuals with a more deliberative cognitive style are more likely to choose options that increase the counterparts' payoffs at a *very low* cost for the decision maker thus promoting social efficiency whereas a more intuitive cognitive style is related to choices that either equalize payoffs between themselves and others (i.e. egalitarian choices) or maximize their own payoff relative to their counterparts (i.e. spiteful choices) (12, 29). In sum, the trait-level analysis largely supports our hypothesis that deliberation favours social efficiency by overriding the individuals' intuitive tendency to care for the relative share each person is allocated with.

*Response Times Manipulation and Social Motives*
Panel B in Figures 1 to 4 displays the social motive classification for each experimental condition (time pressure and time delay; US: n=97 and n=87; India: n=63 and n=69, respectively) for both the US and the India samples. The results of the regression analysis are shown in Panel B of Tables S1 to S4. We observe that the direction of the effect of the time condition is the same across countries except for the case of self-interest. The effect of time delay (vs. time pressure) is significantly positive for both social efficiency variables (both p's<0.01; see Panel B in Table S1). In the case of egalitarianism and spitefulness, the effect of time delay is negative and significant for the model-based egalitarian and choice-based spiteful definitions (both p's<0.01). This effect is also negative for the choice-based egalitarian and model-based spiteful definitions but not significant (both p's>0.31). The time manipulation does not exert a significant effect on self-interest (p=0.83). As shown in Tables S5 to S8 (panel B), the interaction between condition and country is never significant (all p's>0.19).

Subjects' level of experience in similar experiments has been shown to moderate the effect of response time manipulations on behaviour in social dilemmas. Experienced subjects are typically less responsive to manipulations in games they have been previously exposed to (42, 45, 46). To account for this well-documented effect, we decided to provide a robustness check for our findings by restricting the analysis to inexperienced subjects (n=100). We find that the effect of time delay on self-interest becomes similar across countries (see Panel C in Figures 1 to 4). In this inexperienced sample, time delay exerts a marginally significant positive effect on self-interest (p=0.06, Panel C in Table S4), whereas the rest of the results remain qualitatively unaffected (panel C in Tables S1 to S3) except for choice-based social efficiency, which loses its significance (p=0.17). The interaction terms between condition and country keep being non-significant (p's>0.36; see Tables S5 to S8, panel C) except for choice-based social efficiency (p=0.06). A Wald test reveals that the effect of time delay on choice-based social efficiency is significantly positive for the US sample (p=0.03) but non-significant for the India sample (p=0.68).

Thus, at the state level of analysis, the results are also consistent with our hypothesis that deliberation increases concerns for social efficiency by overriding individuals' intuitive tendency to focus on their relative shares.

Regarding differences between countries, residents in India are more likely than residents in the US to be classified as spiteful (p's<0.01 in both the whole and the inexperienced sample) and less likely to favour social efficiency (except for the model-based definition in the inexperienced sample, p=0.22, the country variable is significant in all cases, p's<0.05). This is also in line with the results previously described.

## Discussion

Across two different countries and at both the trait and the state levels of analysis, we found strong evidence that: (i) intuition promotes individuals' concern for relative payoffs (egalitarian and spiteful choices) and (ii) deliberation promotes individuals' concern for social efficiency. Our results suggest that, as hypothesized, deliberation favours social efficiency by overriding the intuitive tendency of individuals to be driven by distributive concerns.

Moreover, the qualitative nature of our main findings does not crucially depend on whether we use a "model-based" or a "choice-based" classification of subjects. While it is true that non-significant effects of deliberation vs. intuition are observed for one of the two definitions in some cases, the effects at either the trait or the state level (even when considering each country separately) never contradict our hypothesized relationships between deliberation and social motives. Additionally, our arguments are also robust to analysing each decision separately (see Tables S12 and S13, and the discussion there). One social motive which is intimately linked to, and can be confounded with, the notion of social efficiency is the Rawlsian *maximin* preference (1, 8, 14), according to which individuals wish to maximize the payoff of the less well-off individual in the group. As shown in the SI, however, the effects observed when analysing each decision separately do not support the existence of a relationship between deliberation/intuition and maximin preferences.

The evidence presented here suggests that a substantial proportion of individuals care about *both* the total surplus and their relative shares. Yet, these social preferences do not occur simultaneously since egalitarian and spiteful concerns seem to be automatic with further deliberation leading people to override them in favour of social efficiency. This also means that a dual-process approach cannot help us pinpoint the main drivers of the difference between egalitarian and spiteful motives.

Our findings are particularly interesting since the relationship between group resources and the way they are to be shared has been a continuing source of debate within distributive justice and social choice theory (1-4). Our data suggest that people's reliance on either intuitive or deliberative decision making affects the extent to which distributive or efficiency concerns dominate. These results are consistent with previous research showing that deliberation favours utilitarian judgments in moral dilemmas (24-28), that equality concerns are rooted in intuitive emotional processing (4, 22, 23), and that fairness is intuitive (47, 48). Our evidence qualifies previous findings by showing that it is not only egalitarianism *per se* but, more generally, the concern for individuals' relative payoffs that responds to intuition.

In addition, we find some indication that deliberation (high CRT scores and time delay – among inexperienced subjects only) may lead to more self-interested decision making. This result is in line with previous research suggesting that deliberation makes people pursue strategies that maximize their material payoffs (45, 49-52). However, this result does not hold in the state-level analysis using the whole sample (both inexperienced and experienced subjects), which may have been due to the fact that experience blurs the effect of cognitive manipulations (42, 45, 46, 53, 54). Understanding whether deliberation promotes self-interested choices and the extent to which previous experience moderates these effects are important questions for future research.

Note also here that both social efficiency and self-interest relate with absolute payoffs (for the group and the self, respectively). Thus, an interpretation of our findings might be that people make relative comparisons intuitively but need deliberation to focus on, or process information from, absolute values. One may speculate that the information associated to any outcome has to be transformed into relative values in order to be processed. This could imply that comparison values are processed automatically whereas absolute values are not. Given the often suggested link between emotion and intuitive processing (17, 45), one possibility is that attribute comparisons are more emotionally charged than absolute attributes and this is why they are processed more automatically. Disregarding relative comparisons may therefore require inhibiting an emotional response. Future research should explore the validity of these arguments in greater detail, within and beyond the social domain.

Related experiments on one-shot social dilemmas suggest that the decision to cooperate is intuitive whereas further deliberation leads individuals to free-ride on the efforts of others (45, 49-52, 55). However, although cooperation is socially efficient in social dilemmas, the decision to cooperate could also stem from egalitarian and reciprocal concerns depending on the players' expectations about others' behaviour. In addition, free-riding is socially inefficient but can result from self-interested, egalitarian, spiteful or reciprocal motives (5, 8, 12, 13). Thus, if social efficiency concerns (and probably self-interest) require deliberation while egalitarian and spiteful motives, as well as reciprocity (54), respond to intuition, the net effect of promoting intuition vs. deliberation on social dilemma behaviour is not straightforward. This could partially explain why a number of studies have failed to find consistent effects or have even yielded conflicting results (46, 56-60).

Regarding differences between countries, we have shown that Indians are in general more likely than Americans to be classified as spiteful and less likely to be classified as socially efficient. These results are consistent with previous research suggesting that India residents are more spiteful (43), less cooperative (42), and less altruistic (61) than US residents. In addition, this observation adds support to the robustness of our main findings since the observed effects (both at the trait and state level) are remarkably similar across countries, regardless of being two societies with seemingly different social preferences at the aggregate level.

Moreover, the differences observed between our experimental treatments indicate that individuals' social motives can be, at least partially, exogenously manipulated. This may have important implications for the design of mechanisms and institutions aimed at promoting certain social or behavioural outcomes. Future state-level investigations should also go beyond time constraints. The use of time constraints, instead of other cognitive manipulations (such as cognitive load, ego-depletion, or conceptual priming), was motivated by the observation that many social and economic interactions require people to make decisions as quickly as possible. Traders and last-minute bidders, for example, have to make decisions within seconds after new information is acquired (62-64). Also, social interactions often require quick decision making, for example, because deliberating may be met with distrust by observers (65-68). However, many social and economic interactions also occur when people are hungry or thirsty, or when they have experienced fatigue, suggesting that cognitive load or ego depletion are particularly relevant manipulations. Since these factors have been shown to impair deliberative processing and affect behaviour in a number of situations (69-72), it would be fruitful to extend our analysis to these other cognitive manipulations.

Finally, in this study and for the sake of focusing on the conflict between total and relative payoffs, we have analysed social efficiency, egalitarian and spiteful motives. Indeed, previous research emphasizes the relevance of this categorization (5, 8, 10). However, other social motives have been considered in the literature, such as hyper-altruism (i.e. weighting the other's payoff more than one's own (73-75)) and extreme altruism (risking one's own life to save someone else's (76)). Further research may use a different set of decision problems to classify these other motives.

# Methods

*Design and procedure*
We conducted the experiments with participants from the US and India using monetary incentives. The stakes for the experiment conducted with Indian participants were one third of the stakes in the experiment with US participants (expressed in Indian Rupees and US dollars, respectively). This was done to equate the purchasing power of participants' payments in both countries according to the latest data from the World Bank (http://data.worldbank.org/indicator/PA.NUS.PPPC.RF). Since the two studies differed only with regards to the monetary incentives, we report here only the details about the experimental procedure used with the US subjects.

Subjects were recruited using Amazon Mechanical Turk (77-81) and earned $0.90 for participating in a 15-minute (mean=23, median=16) study. In addition, they received an extra payment depending on their performance during the experiment. Participation fees and extra payments were chosen in order to guarantee that participants, on average, would earn a total bonus above the minimum wage of the country where they were based at the time of the experiment. Although stakes were well below the average stake used in standard laboratory experiments, they were within the range of AMT experiments. A number of studies have shown that data collected using AMT are both qualitatively and quantitatively comparable with those collected using the standard physical laboratory (77-81). Moreover, several studies have investigated the effect of stakes on pro-sociality and found that pro-social choices decrease when passing from no-stakes to small-stakes (82), but then they are stake-independent (as long as stakes are not too high) in a number of economic games including the Dictator Game (83), the Public Goods game (84), the Trust Game (85), and the Ultimatum Game (86). A recent study even shows that well-established patterns of social behaviour which are observed in complex interactive experiments in the laboratory are replicated in AMT (87). In sum, previous research supports the claim that data gathered using AMT with relatively small stakes are of no less quality than those collected in the standard physical laboratory.

After entering their MTurk ID, participants were randomly assigned to one of three conditions: *neutral*, *time pressure* or *time delay*. In each condition, participants were asked to make six binary decisions about how to allocate a number of points (10 points = $0.90) between themselves and another anonymous participant they were matched with. These decision problems were used to infer individuals' social motives, as in Study 2 of Corgnet et al. (12). In the *time pressure* condition, participants were asked to make each choice within 5 seconds. In the *time delay* condition, they were asked to wait for at least 15 seconds before making each choice. The time limits (<5s vs. >15s) were chosen following previous research (58). Subjects' average response time was 2.14 and 22.57 seconds in the time pressure and time delay conditions, respectively (this difference is significant, $p<0.01$, t-test). Only subjects who respected the time constraints are considered for the analyses. If we include those subjects who did not respect them (56), the results are qualitatively similar (see Tables S14 to S17). In the *neutral* condition, participants were left free to make their choices at any time (average response time = 5.40 seconds, which differs significantly from

the other two conditions, both p's<0.01, t-test). See the next subsection for the exact decision problems.

After the decisions were made, we asked four comprehension questions. Subjects failing any comprehension question were automatically excluded from the experiment and received no payment.

Subjects who passed the comprehension questions then completed a Numeracy Test (88, 89) and an extended 7-item Cognitive Reflection Test (32, 33). We included the Numeracy Test to assess whether any relation between CRT scores and choices could be due to computational skills rather than to one's capacity to reflect/deliberate (33-35). Controlling for numeracy in our analysis is essential because solving CRT questions not only requires blocking incorrect intuitive answers but it also entails basic computation skills to find the correct answer to the problem. Indeed, scores in both tests are highly correlated (Spearman $\rho=0.60$, $p<0.01$, $n=192$). We modified the original CRT questions in (32) and (33) so that Mturkers could not access the answers online while completing the study, which may be a serious issue (90). We thus changed the context and the numerical solutions of the original CRT questions without changing the spirit of the test. The CRT was included at the end of the experiment to avoid priming reflective processing (27) thus distorting the relationship between social behaviour and reflection. Correct answers were incentivized with a $0.06 reward. As is standard, no time restriction was imposed in any of the tests. Both tests can be found in the Supplementary Information (SI).

Finally, subjects filled a questionnaire with the usual socio-demographic questions. To analyse the role of experience (42, 45, 46, 49, 53, 54), we asked subjects "To what extent have you previously participated in other studies like this one (i.e., exchanging money with a stranger)?". Responses were collected using a 5-point likert-scale from 1 = "Never" to 5 = "Several times". See the SI for full experimental instructions.

*Social motives elicitation*
In each decision problem, participants were asked to choose between the egalitarian Option A and the non-egalitarian Option B: Option A always allocates 10 points to the decision maker (DM) and 10 points to the other participant, whereas the distribution of points associated with Option B depends on the decision problem (see Table 1).

Participants were informed that their final payoff would be determined by only one decision selected at random. In this way, we encouraged participants to treat each decision independently.

This task is particularly suited to analyse the cognitive underpinnings of social behaviour because it is short and cognitively undemanding (12). In addition, it allows us to assess possible asymmetries in social preferences related to either advantageous or disadvantageous payoff comparisons (9). Thus, the task provides a good balance between the amount of information gathered and the complexity of the decisions. We classify individuals' choices as follows:

i. Socially *efficient*, if they maximize the total joint payoff;
ii. *Egalitarian*, if they minimize payoff inequality;
iii. *Spiteful*, if they maximize the decision maker's relative standing by minimizing the other's payoff;
iv. *Self-interested*, if they maximize the decision maker's own payoff.

Importantly, we do not force a trade-off between any two types of motives across decisions but it is instead an individual's complete set of choices that allows us to infer her motives. In some decisions in our task, for instance, there is a conflict between egalitarian and socially efficient options, whereas in others equality and social efficiency are aligned but in conflict with self-interest and/or spitefulness. Table 1 shows the motives that are consistent with each option in each decision. It can be seen that from one single decision it would be hard to say with certainty which social motive is driving choice. This happens in nearly all economic games on social preferences (8, 12). Therefore, we need to analyse the consistency of motives across decisions.

Table 1. Association between choices and social preferences.

| Decision # | Option A payoffs (DM, other) | Option B payoffs (DM, other) | Motives consistent with Option A | Motives consistent with Option B |
|---|---|---|---|---|
| Decision 1 | (10, 10) | (10, 6) | Efficiency<br>Egalitarian<br>Self-interest | Spiteful<br>Self-interest |
| Decision 2 | (10, 10) | (16, 4) | Efficiency<br>Egalitarian | Efficiency<br>Spiteful<br>Self-interest |
| Decision 3 | (10, 10) | (10, 18) | Egalitarian<br>Spiteful<br>Self-interest | Efficiency<br>Self-interest |
| Decision 4 | (10, 10) | (11, 19) | Egalitarian<br>Spiteful | Efficiency<br>Self-interest |
| Decision 5 | (10, 10) | (12, 4) | Efficiency<br>Egalitarian | Spiteful<br>Self-interest |
| Decision 6 | (10, 10) | (8, 16) | Egalitarian<br>Self-interest<br>Spiteful | Efficiency |

Notes: One motive is associated to both options when an individual motivated by such a preference would be indifferent between Options A and B in that decision.

## *Statistical analysis*

For each of the three social motives we consider two alternative definitions. First, we classify subjects using a generalized Fehr-Schmidt (9) model which is extensively used in social preferences research and has been used in previous studies (12, 29). The "model-based" definition captures those subjects whose choices are perfectly consistent with the parameters of a *generalized* Fehr-Schmidt (9) model characterizing a particular motive (12) (see SI). Alternative approaches such as the Charness-Rabin (8) model would result in an identical classification of subjects, since the parameters of the basic Charness-Rabin model are linear transformations of the Fehr-Schmidt parameters (8, 91). Second, following the tradition of research on social value orientation (10), we also consider a "choice-based" definition in which at least 2/3 of the choices (i.e. 4 or more) are consistent with that specific motive. We obtain similar results if we use a "3 or more" or a "5 or more" criterion instead.  Model-based definitions to classify people follow the standards in economics whereas social psychology research has traditionally relied on choice-based definitions. Since both types of definitions have their own (dis)advantages and we do not find any reason to favour one of these two research traditions over the other, we show the results for both definitions. In fact, this strengthens our findings. The Spearman correlations between the two definitions are 0.41, 0.60 and 0.44 (all p's<0.01, n=508) for efficiency, egalitarian and spiteful motives, respectively. The classification of subjects according to the model-based definitions leads to mutually exclusive categories; however, this is not the case for the choice-based definitions. Note that both definitions are equivalent for self-interest.

In our analyses, we exclude those subjects (13%) whose choices were inconsistent (i.e. the subject chose to increase/reduce the counterpart's payoff in one decision but s/he did not take the same action in another decision where doing so was less costly), which also means that we can obtain a reliable range for the parameters used in the model-based definition for all subjects included.

## Ethics statement

All subjects provided written informed consent prior to participating. This research was conducted while the first author was affiliated to the Center for Mathematics and Computer Science (CWI), Amsterdam. According to the Dutch legislation, this is a non-WMO study which (i) does not involve medical research and (ii) participants are not asked to follow rules of behaviour. See http://www.ccmo.nl/attachments/files/wmo-engelse-vertaling-29-7-2013-afkomstig-van-vws.pdf, §1, Article 1b, for an English translation of the Medical Research Act. Thus (see http://www.ccmo.nl/en/non-wmo-research) the only legislations which apply are the Agreement on Medical Treatment Act, from the Dutch Civil Code (Book 7, title 7, §5), and the Personal Data Protection Act (a link to which can be found in the previous webpage). The current study conforms to both.

## Data accessibility

All data can be accessed at doi:10.5061/dryad.n581t.

## Competing interests

The authors declare that they have no competing interests.

## Authors' contributions

V.C., B.C., A.M.E., and R.H.G. designed the experiment, conducted the experiment, analysed the data, and wrote the manuscript.

**Acknowledgments**

The authors acknowledge general financial support from the Economic Science Institute, the International Foundation for Research in Experimental Economics, the Argyros School of Business and Economics at Chapman University, the University of Nottingham Business School, the Spanish Ministry of Education [Grant 2012/00103/001], Ministry of Economy and Competence [2016/00122/001], Spanish Plan Nacional I+D MCI [ECO2013-44879-R], 2014-17, and Proyectos de Excelencia de la Junta Andalucía [P12.SEJ.1436], 2014-18.

# References

1. Rawls J. 1971 *A theory of justice*. Cambridge, MA: Harvard University Press.
2. Sen AK. 1984 *Collective choice and social welfare*. New York, NY: Elsevier.
3. Kohlberg L. 1981 *The philosophy of moral development: Moral stages and the idea of justice.* San Francisco, CA: Harper and Row.
4. Hsu M, Anen C, Quartz SR. 2008 The Right and the Good: Distributive justice and neural encoding of equity and efficiency. *Science*, 320, 1092-1095.
5. Fehr E, Schmidt KM. (2006) The Economics of Fairness, Reciprocity and Altruism - Experimental Evidence and New Theories. In: Kolm, S.C. (ed.): *Handbook of the Economics of Giving, Altruism and Reciprocity. Handbooks in Economics* 23, Vol. 1. Amsterdam, NE: Elsevier.
6. Brañas-Garza P, Espín AM, Exadaktylos F, Herrmann B. 2014 Fair and unfair punishers coexist in the Ultimatum Game. *Sci. Rep.* 4, 6025.
7. Espín AM, Exadaktylos F, Herrmann B, Brañas-Garza P. 2015 Short-and long-run goals in ultimatum bargaining: impatience predicts spite-based behavior. *Front. Behav. Neurosci.* 9, 214.
8. Charness G, Rabin M. 2002 Understanding social preferences with simple tests. *Q. J. Econ.* 117, 817-869.
9. Fehr E, Schmidt KM. 1999 A theory of fairness, competition, and cooperation. *Q. J. Econ.* 114, 817-868.
10. Van Lange PAM, Otten W, De Bruin EMN, Joireman JA. 1997 Development of prosocial, individualistic, and competitive orientations: Theory and preliminary evidence. *J. Person. Soc. Psychol.* 73, 733-746.
11. Messick DM, McClintock CG. 1968 Motivational bases of choice in experimental games. *J. Exp. Soc. Psychol.* 4, 1-25.
12. Corgnet B, Espín AM, Hernán-González R. 2015 The cognitive basis of social behavior: cognitive reflection overrides antisocial but not always prosocial motives. *Front. Behav. Neurosci.* 9, 287.

13. Espín AM, Brañas-Garza P, Herrmann B, Gamella JF. 2012 Patient and impatient punishers of free-riders. *Proc. R. Soc. B* 279, 4923-4928.
14. Engelmann D, Strobel M. 2004 Inequality aversion, efficiency, and maximin preferences in simple distribution experiments. *Am. Econ. Rev.* 94, 857-869.
15. Capraro V. 2013 A model of human cooperation in social dilemmas. *PLoS ONE* 8, e72427.
16. Chaiken S, Trope Y. 1999 *Dual-process theories in social psychology*. New York, NY: Guilford press.
17. Kahneman D. 2011 *Thinking, fast and slow*. New York, NY: Farrar, Straus, & Giroux.
18. Sloman S. 1996 The empirical case for two systems of reasoning. *Psychol. Bull.* 119, 3-22.
19. Evans JSB. 2003 In two minds: dual-process accounts of reasoning. *Trends Cogn. Sci.* 7, 454-459.
20. Krajbich I, Bartling B, Hare T, Fehr E. 2015 Rethinking fast and slow based on a critique of reaction-time reverse inference. *Nat. Commun.* 6, 7455.
21. Kraft-Todd G, Yoeli E, Bhanot S, Rand DG. 2015 Promoting cooperation in the field. *Curr. Opin. Behav. Sci.* 3, 96-101.
22. Roch SG, Lane JA, Samuelson CD, Allison ST, Dent JL. 2000 Cognitive load and the equality heuristic: a two-stage model of resource overconsumption in small groups. *Organ. Behav. Hum. Decis. Process.* 83, 185-212.
23. Feng C, Long Y-J, Krueger F. 2015 Neural signatures of fairness-related normative decision making in the ultimatum game: A coordinate-based meta-analysis. *Hum. Brain Mapp.* 36, 591-602.
24. Conway P, Gawronski B. 2013 Deontological and utilitarian inclinations in moral decision making: A process dissociation approach. *J. Pers. Soc. Psychol.* 104, 216-235.
25. Greene JD, Morelli SA, Lowenberg K, Nystrom LE, Cohen JD. 2008 Cognitive load selectively interferes with utilitarian moral judgment. *Cognition* 107, 1144-1154.
26. Suter RS, Hertwig R. 2011 Time and moral judgment. *Cognition* 119, 295-300.
27. Paxton JM, Ungar L, Greene JD. 2012 Reflection and reasoning in moral judgment. *Cogn. Sci.* 36, 163-177.
28. Trémolière B, De Neys W, Bonnefon J-F. 2012 Mortality salience and morality: Thinking about death makes people less utilitarian. *Cognition* 124, 379-384.
29. Ponti G, Rodriguez-Lara I. 2015 Social preferences and cognitive reflection: evidence from a dictator game experiment. *Front. Behav. Neurosci.* 9, 146.
30. Sheremeta RM. 2015 Impulsive behavior in competition: Testing theories of overbidding in rent-seeking contests. *Available at* http://ssrn.com/abstract=2676419.
31. Bartling B, Fehr E, Maréchal MA, Schunk D. 2009 Egalitarianism and competitiveness. *Am. Econ. Rev.* 99, 93-98.
32. Frederick S. 2005 Cognitive reflection and decision making. *J. Econ. Perspect.* 19, 25-42.
33. Toplak ME, West RF, Stanovich KE. 2014 Assessing Miserly information processing: An expansion of the cognitive reflection test. *Think. Reason.* 20, 147-168.

34. Sinayev A, Peters E. 2015 Cognitive reflection vs. calculation in decision making. *Front. Psychol.* 6, 532.
35. Thomson KS, Oppenheimer DM. 2016 Investigating an alternate form of the cognitive reflection test. *Judgm. Decis. Mak.* 11, 99-113.
36. Shiffrin RM, Schneider W. 1977 Controlled and automatic information processing: II. Perceptual learning, automatic attending, and a general theory. *Psychol. Rev.* 84, 127-190.
37. Miller EK, Cohen JD. 2001 An integrative theory of prefrontal cortex function. *Ann Rev. Neurosci.* 24, 167-202.
38. Gächter S, Schulz JF. 2016 Intrinsic honesty and the prevalence of rule violations across societies. *Nature* 531, 496-499.
39. Stagnaro MN, Arechar AA, Rand DG. 2016 From good institutions to good norms: Tod-down incentives to cooperate foster prosociality but not norm enforcement. *Available at* http://ssrn.com/abstract=2720585.
40. Guha R. 2007 *India after Gandhi: The history of the world's largest democracy*. New York, NY: HarperCollins.
41. Quah JS. 2008 Curbing corruption in India: An impossible dream? *Asian J Polit Sci* 16, 240-259.
42. Capraro V, Cococcioni G. 2015 Social setting, intuition, and experience in laboratory experiments interact to shape cooperative decision-making. *Proc. R. Soc. B* 282, 20150237.
43. Fehr E, Hoff K, Kshetramade M. 2008 Spite and development. *Am. Econ. Rev.* 98, 494-499.
44. Bosch-Domènech A, Brañas-Garza P, Espín AM. 2014 Can exposure to prenatal sex hormones (2D: 4D) predict cognitive reflection? *Psychoneuroendocrinology 43*, 1-10.
45. Rand DG, Peysakhovich A, Kraft-Todd GT, Newman GE, Wurzbacher O, Greene JD, Nowak MA. 2014 Social heuristics shape intuitive cooperation. *Nat Commu* 5, 3677.
46. Capraro V, Cococcioni G. 2016 Rethinking spontaneous giving: Extreme time pressure and ego-depletion favor self-regarding reactions. *Sci. Rep. 6, 27219*.
47. Cappelen AW, Nielsen UH, Tungodden B, Tyran J-R, Wengström E. In press. Fairness is intuitive. *Exp. Econ*.
48. Artavia-Mora L, Bedi AS, Rieger M. 2016 Intuitive cooperation in the field. *Available at SSRN: http://www.ssrn.com/abstract=2679179*.
49. Rand DG, Greene JD, Nowak MA. 2012 Spontaneous giving and calculated greed. *Nature* 489, 427-430.
50. Bear A, Rand DG. 2016 Intuition, deliberation, and the evolution of cooperation. *Proc Natl Acad Sci USA* 113, 936-941.
51. Rand DG, Brescoll VL, Everett JAC, Capraro V, Barcelo H. 2016 Social heuristics and social norms: Intuition favors altruism for women but not for men. *J. Exp. Psychol. Gen.* 145, 389-396.
52. Rand DG. 2016 Cooperation, fast and slow: Meta-analytic evidence for a theory of social heuristics and self-interested deliberation. *Psychol. Sci*. doi: 10.1177/0956797616654455
53. Chandler J, Paolacci G, Peer E, Mueller P, Ratliff KA. 2015 Using nonnaive participants can reduce effect sizes. *Psychol. Sci.* 26, 1131-1139.

54. Rand DG, Kraft-Todd GT. 2014 Reflection does not undermine self-interested prosociality. *Front. Behav. Neurosci.* 8, 300.
55. Rand DG, Nowak MA. 2013 Human cooperation. *Trends Cogn Sci* 17, 413-425.
56. Tinghög G, *et al.* 2013 Intuition and cooperation reconsidered. *Nature*, 498, E1-E2.
57. Verkoeijen PP, Bouwmeester S. 2014 Does intuition cause cooperation?. *PloS One* 9, e96654.
58. Myrseth KOR, Wollbrant C. 2015 Intuitive cooperation Refuted: Commentary on Rand et al. (2012) and Rand et al. (2014). *University of Gothenburg, Working Papers in Economics* No. 617.
59. Lohse J. In press. Smart or selfish – When smart guys finish nice. *J. Behav. Exp. Econ.*
60. Dickinson DL, McElroy T. 2016 Moderate sleep restriction and time-of-day impacts on simple social interactions. *Available at* http://www.chapman.edu/research-and-institutions/economic-science-institute/_files/ifree-papers-and-photos/Sleep-Barg-Trust-Dickinson-April-2016.pdf.
61. Raihani NJ, Mace R, Lamba S. 2013 The effect of $1, $5 and $10 stakes in an online Dictator Game. *PLoS ONE* 8, e73131.
62. Busse JA, Green C. 2002 Market efficiency in real time. *J. Financ. Econ.* 65, 415-437.
63. Kocher MG, Pahlke J, Trautmann ST. 2013 Tempus fugit: time pressure in risky decisions. *Manag. Sci.* 59, 2380-2391.
64. Roth AE, Ockenfels A. 2002 Last-minute bidding and the rules for ending second-price auctions: Evidence from eBay and Amazon auctions on the Internet. *Am. Econ. Rev.* 92, 1138-1151.
65. Hilbe C, Hoffman M, Nowak MA. 2015 Cooperate without looking in a non-repeated game. *Games* 6, 458-472.
66. Hoffman M, Yoeli E, Nowak MA. 2015 Cooperate without looking: Why we care what people think and not just what they do. *Proc. Natl. Acad. Sci.* 112, 1727-1732.
67. Capraro V, Kuilder J. 2015 To know or not to know? Looking at payoffs signals selfish behavior but it does not actually mean so. *Available at* http://ssrn.com/abstract=2679326.
68. Jordan JJ, Hoffman M, Nowak MA, Rand DG. 2016 Uncalculating cooperation as a signal of trustworthiness. *Proc. Natl. Acad. Sci.* doi: 10.1073/pnas.1601280113
69. Baumeister RF, Bratslavsky E, Muraven M, Tice DM. 1998 Ego depletion: Is the active self a limited resource? *J. Pers. Soc. Psychol.* 74, 1252-1265.
70. Mead NL, Baumeister RF, Gino F, Schweitzer ME, Ariely D. 2009 Too tired to tell the truth: Self-control resource depletion and dishonesty. *J. Exp. Soc. Psychol.* 45, 594-597.
71. Xu H, Bègue L, Bushman, BJ. 2012 Too fatigued to care: Ego depletion, guilt, and prosocial behavior. *J. Exp. Soc. Psychol.* 48, 1183-1186.
72. Yam KC, Reynolds SJ, Hirsch JB. 2014 The hungry thief: Physiological deprivation and its effects on unethical behavior. *Organ. Behav. Hum. Decis. Process.* 125, 123-133.
73. Capraro V. 2015 The emergence of hyper-altruistic behaviour in conflictual situations. *Sci. Rep.* 4, 9916.

74. Crockett MJ, Kurth-Nelson Z, Siegel JZ, Dayan P, Dolan RJ. 2014 Harm to others outweighs harm to self in moral decision making. *Proc. Natl. Acad. Sci.* 111, 17320-17325.
75. Kitcher P. 1993 The evolution of human altruism. *J. Philos.*, 90, 497-516.
76. Rand DG, Epstein ZG. 2014 Risking your life without a second thought: Intuitive decision-making and extreme altruism. *PLoS ONE* 9, e109687.
77. Paolacci G, Chandler J, Ipeirotis PG. 2010 Running experiments on Amazon Mechanical Turk. *Judgm. Dec. Making* 5, 411-419.
78. Horton JJ, Rand DG, Zeckhauser RJ. 2011 The online laboratory: Conducting experiments in a real labor market. *Exp. Econ.* 14, 399-425.
79. Rand DG. 2012 The promise of Mechanical Turk: How online labor markets can help theorists run behavioral experiments. *J. Theor. Biol.* 299, 172-179.
80. Berinsky AJ, Huber GA, Lenz GS. 2012 Evaluating online labor markets for experimental research: Amazon.com's Mechanical Turk. *Polit. Anal.* 20, 351-368.
81. Paolacci G, Chandler J. 2014 Inside the Turk: Understanding Mechanical Turk as a participation pool. *Curr. Dir. Psychol. Sci.* 23, 184-188.
82. Amir O, Rand DG, Gal YK. 2012 Economic games on the Internet: The effect of $1 stakes. *PLoS ONE* 7, e31461.
83. Forsythe R, Horowitz JL, Savin NE, Sefton M. 1994 Fairness in simple bargaining experiments. *Games Econ. Behav.* 6, 347-369.
84. Kocher MG, Martinsson P, *Visser M.* 2008 Does stake size matter for cooperation and punishment? *Econ. Lett.* 99, 508-511.
85. Johansson-Stenman O, Mahmud M, Martinsson *P.* 2005 Does stake size matter in trust games? *Econ. Lett.* 88, 365-369.
86. Cameron LA. 1999 Raising the stakes in the ultimatum game: Experimental evidence from Indonesia. *Econ. Inq.* 37, 47-59.
87. Arechar AA, Gächter S, Molleman L. 2016 Conducting interactive experiments online. *Available at SSRN: https://ssrn.com/abstract=2884409.*
88. Schwartz LM, Woloshin S, Black WC, Welch HG. 1997 The role of numeracy in understanding the benefit of screening mammography. *Ann. Intern. Med.* 127, 966-972.
89. Cokely ET, Galesic M, Schulz E, Ghazal S, Garcia-Retamero R. 2012 Measuring risk literacy: The Berlin Numeracy Test. *Judgm. Dec. Making* 7, 25-47.
90. Brañas-Garza P, Kujal P, Lenkei B. 2015 Cognitive Reflection Test: Whom, how, when. *MPRA Paper* 68049.

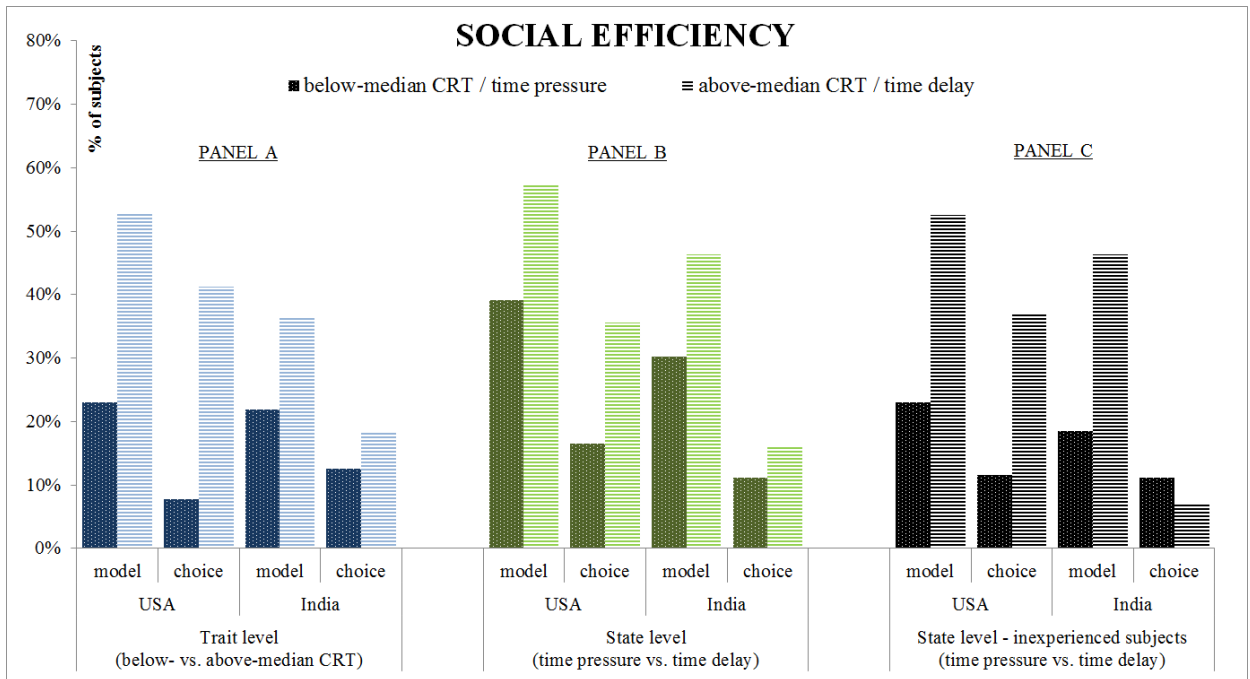

*Figure 1.* Proportion of subjects classified as Socially Efficient, broken down into below- and above-median CRT scores (panel A; below/above-median CRT: n=65/51 in US, n=32/44 in India), time pressure and time delay for all subjects (panel B; time pressure/delay: n= 97/87 in US, n= 63/69 in India) and for inexperienced subjects only (panel C; time pressure/delay: n= 26/19 in US, n= 27/28 in India).

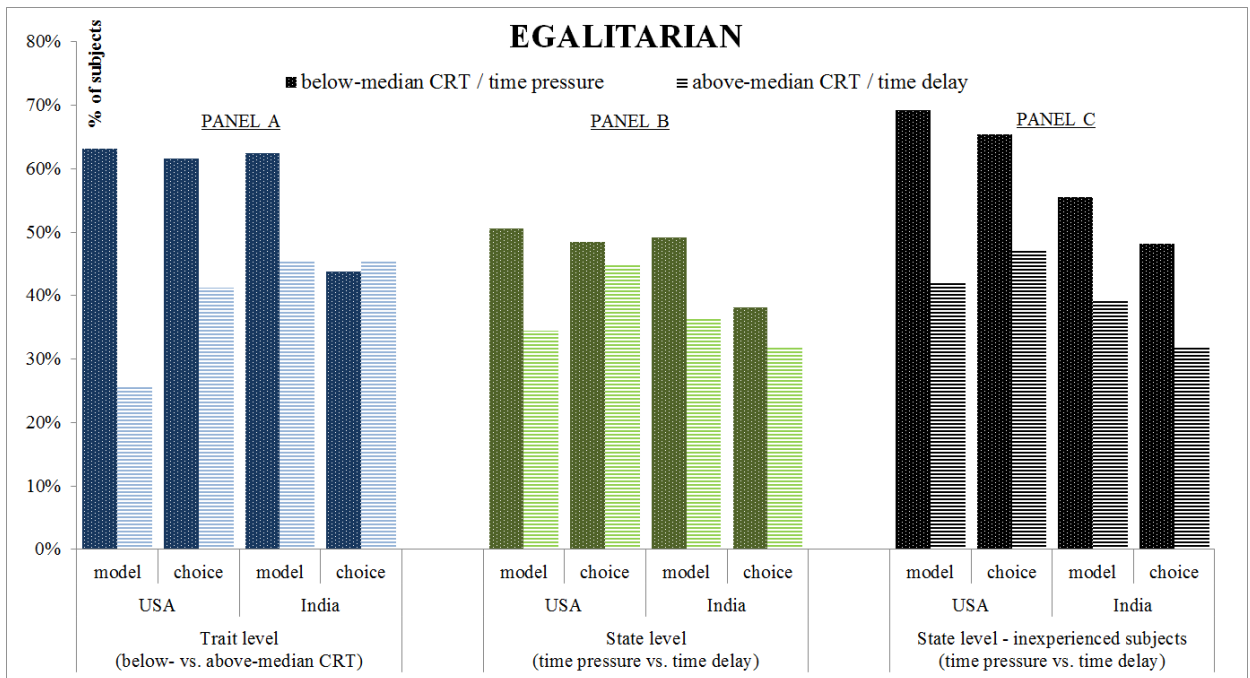

*Figure 2.* Proportion of subjects classified as Egalitarian, broken down into below- and above-median CRT scores (panel A), time pressure and time delay for all subjects (panel B) and for inexperienced subjects only (panel C). See caption of Figure 1 for the number of observations in each subgroup.

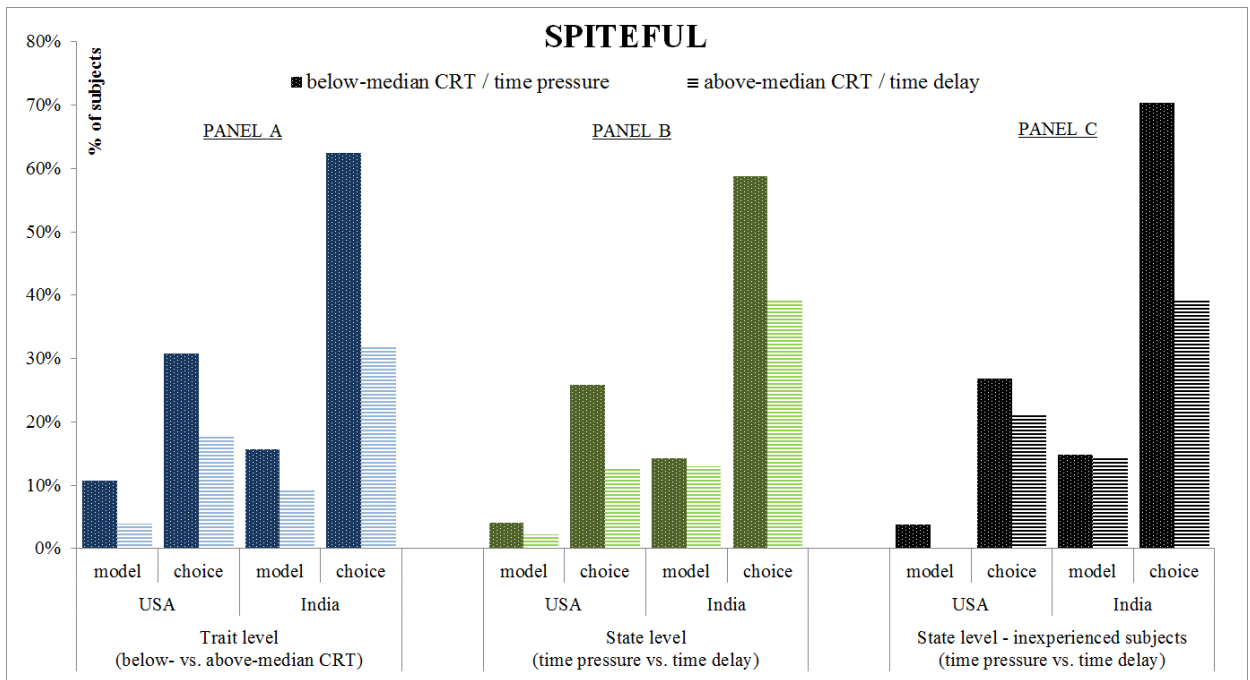

*Figure 3.* Proportion of subjects classified as Spiteful, broken down into below- and above-median CRT scores (panel A), time pressure and time delay for all subjects (panel B) and for inexperienced subjects only (panel C). See caption of Figure 1 for the number of observations in each subgroup.

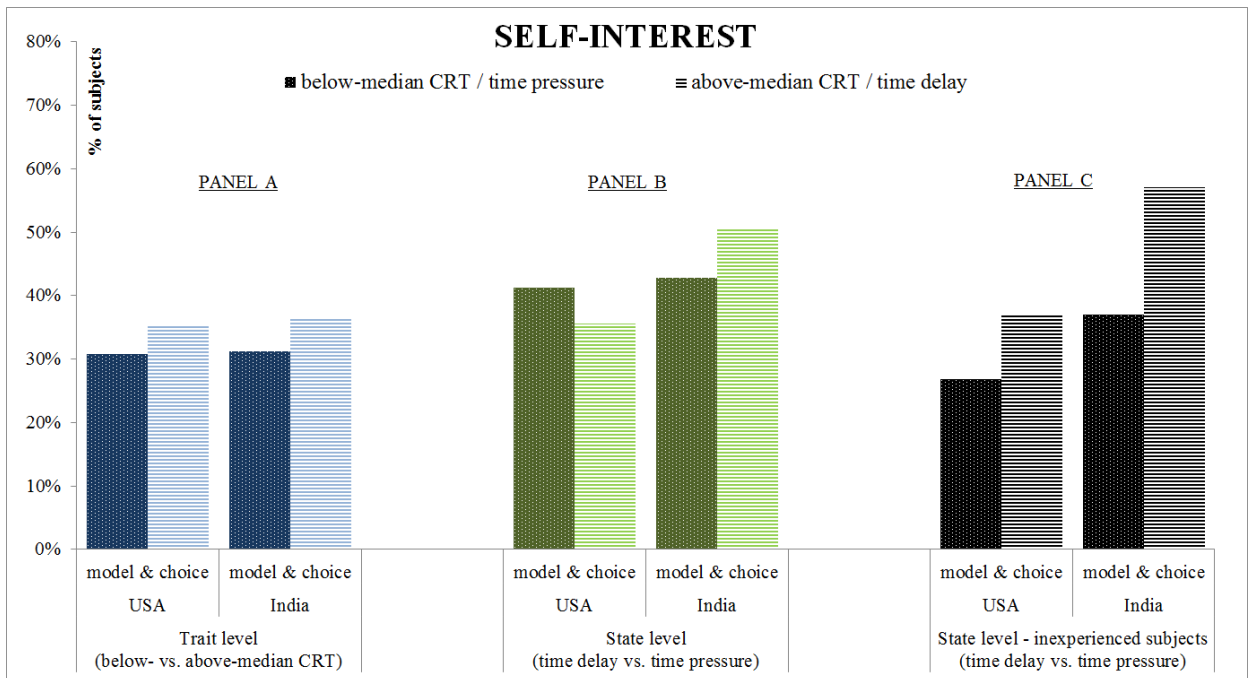

*Figure 4.* Proportion of subjects classified as Self-Interested, broken down into below- and above-median CRT scores (panel A), time pressure and time delay for all subjects (panel B) and for inexperienced subjects only (panel C). See caption of Figure 1 for the number of observations in each subgroup.